\documentclass[12pt, journal, twocolumn]{IEEEtran}
\usepackage{cite}
\usepackage{amsmath,amssymb,amsfonts}
\usepackage{algorithmic}
\usepackage{graphicx}
\usepackage{textcomp}
\usepackage{url}
\usepackage{xcolor, soul}
\usepackage{eurosym}
\usepackage{array}

\makeatletter
\def\thickhline{%
  \noalign{\ifnum0=`}\fi\hrule \@height \thickarrayrulewidth \futurelet
   \reserved@a\@xthickhline}
\def\@xthickhline{\ifx\reserved@a\thickhline
               \vskip\doublerulesep
               \vskip-\thickarrayrulewidth
             \fi
      \ifnum0=`{\fi}}
\makeatother

\newlength{\thickarrayrulewidth}
\usepackage{threeparttable}
\usepackage{booktabs}

\def\BibTeX{{\rm B\kern-.05em{\sc i\kern-.025em b}\kern-.08em
    T\kern-.1667em\lower.7ex\hbox{E}\kern-.125emX}}
\begin{document}

\title{Sustainable RF Wireless Energy Transfer for Massive IoT: enablers and challenges}

\author{\IEEEauthorblockN{Osmel M. Rosabal,~\IEEEmembership{Graduate Student Member,~IEEE,}
			Onel L. A. L\'opez,~\IEEEmembership{Member,~IEEE,}
                Hirley Alves,~\IEEEmembership{Member,~IEEE,}
                Matti Latva-aho,~\IEEEmembership{Senior Member,~IEEE}
		}
		\thanks{All authors are with the Centre for Wireless Communications (CWC), University of Oulu, Finland. Email: firstname.lastname@oulu.fi}
		\thanks{This research has been financially supported by the Academy of Finland (Grant 346208 (6G Flagship) and Grant 348515).}}				
	\maketitle

\begin{abstract}
Reliable energy supply remains a crucial challenge in the Internet of Things (IoT). Although relying on batteries is cost-effective for a few devices, it is neither a scalable nor a sustainable charging solution as the network grows massive. Besides, current energy-saving technologies alone cannot cope, for instance, with the vision of zero-energy devices and the deploy-and-forget paradigm which can unlock a myriad of new use cases. In this context, sustainable radio frequency wireless energy transfer emerges as an attractive solution for efficiently charging the next generation of ultra low power IoT devices. Herein, we highlight that sustainable charging is broader than conventional green charging, as it focuses on balancing economy prosperity and social equity in addition to environmental health. We discuss the economic implications of powering energy transmitters with ambient energy sources, and reveal insights on their optimal deployment. Moreover, we overview different methods for modeling the energy arrival process of ambient energy sources and discuss their application in different use cases. We highlight the potential of integrating sustainable WET with energy harvesting from nearby transmitters and discuss enhancements in energy receiver design. We also illustrate the role of different technologies in enabling sustainable WET and exemplify various use cases. Besides, we reveal insights into low-complexity architectures designed at the energy transmitters. We highlight relevant research challenges and candidate solutions. 
\end{abstract}


\maketitle

\section{Introduction}\label{sec:intro}
\label{sec:introduction}
\IEEEPARstart{T}{he} Internet of Things (IoT) revolution calls for the synergy among environmental health, economic vitality, and social advances of current and future communication networks to promote sustainability \cite{matinmikko2020white}. Currently, millions of IoT devices take care of our well-being, monitor environmental pollution, track assets, and help manage natural resources. As the popularity of IoT applications grows, maintaining an uninterrupted operation of a massive number of battery-powered devices becomes challenging. In fact, due to a small form factor, most IoT devices can only carry a tiny battery whose effective lifetime does not match that of the device's electronics \cite{9536560}. This means that battery replacements can grow faster than the number of connected devices, thus, increasing business operational costs and environmental pollution if the electronic waste is not disposed of correctly. Moreover, the maintenance of IoT devices that operate in remote areas, embedded in civil infrastructure, or medical implants, represents a risky, and possibly costly, operation. That is why companies consider the battery lifetime critical when estimating the maintenance costs of the network infrastructure \cite{9536560}.

Two main research directions to enable (and support the perpetual autonomy of) sustainable IoT networks from the communication perspective are: energy-saving mechanisms and energy repletion \cite{9361680}. Energy-saving mechanisms aim at minimizing the energy needed for completing tasks such as sensing, processing, and communication without significantly degrading the system performance. On the other hand, energy repletion comprises energy harvesting (EH) techniques for recharging the batteries by exploiting ambient or dedicated energy sources, also known as power beacons (PBs).

This work focuses on sustainable radio frequency (RF) wireless energy transfer (WET), hereinafter referred to as WET. It is worth noticing that previous research and development directions within the scope of WET have focused on specific performance optimizations. This has been done, for instance, by deploying multiple PBs, using efficient energy beamforming strategies, and enhancing the design of RF-EH receivers \cite{alves2021wireless,9920166,9658133}. Specifically, the authors in \cite{9920166} optimize the deployment of directional PBs to charge mobile sensors considering the physical space each sensor occupies within the PBs' coverage area. Therein, they devise a device-to-PB assignation strategy based on individual energy demands to minimize the total charging costs. Similarly, the authors in \cite{9658133} optimize the position, antenna orientation, and service time allocation of mobile PBs. Therein, they maximize a function of the per-device harvested energy considering that PBs can move within a restricted area. A similar line of work follows from the works proposed in \cite{10057057,10081057,9708417,9319211} which exploit fully moving/flying capabilities at the PBs---a strategy also referred to as nomadic WET---to dynamically shorten the charging distance. For example, the authors in \cite{10057057} minimize the energy expenditure of a moving PB, i.e., the sum of propulsion energy and transmission energy, to meet the devices' charging demands. Therein, the authors optimize the trajectory and charging scheduling of the PB considering omnidirectional radiation and an arbitrary distribution of the devices in the network. The work in \cite{10081057} studies a similar setup but considers a directional PB that charges different groups of devices at each stopping point. The total charging time is minimized by optimizing the PB's path, antenna orientation, and the position of the stopping points for maximum received power of the sensors within the served group. Notice that additional spatial degrees of freedom can be exploited when using flying PBs due to their inherent deployment flexibility and tridimensional coverage. An example of a flying PB-enabled scenario is studied in \cite{9708417}, where the authors devise the minimum PB's energy consumption that meets the devices' charging demands. To achieve that, the authors jointly optimize the charging trajectory, hovering locations, and charging time for each group of devices. In \cite{9319211}, the authors discuss the main features, potentials, and challenges of WET for powering massive IoT deployments. Specifically, they discuss additional enhancements in WET-enabled networks such as deploying radio stripes, rotor-equipped PBs, intelligent reflective surfaces aided WET, and ultra-low power receivers implementation. Moreover, the authors also raise the limitations of accurate channel state information (CSI) estimation in WET and discuss possible solutions. It is worth mentioning other enhancements that have been already proposed for wireless communications and that can be exploited also for WET, such as flexible antennas \cite{zheng2023flexible} realized via fluid and movable antennas. The former case refers to a software-reconfigurable antenna whose radiating properties can be adjusted via fluidic materials, while the latter refers to a mechanically steerable antenna. Regardless of the implementation, flexible antennas provide the tools for overcoming the channel impairments by dynamically reconfiguring the radiator. Furthermore, the integration of renewable sources into WET-enabled networks has been also proposed in the literature either to support the operation of the PBs \cite{9802635,8664000,8309564,9999300,9770091,9319211} or to aid the WET service for reducing PBs' energy consumption \cite{8742610}. Specifically, \cite{9802635} studies the optimal charging scheduling strategy of a network of PB powered by renewable sources to maximize the number of charged devices in a round, \cite{8664000} focuses on the integration of ambient EH and WET, and \cite{8309564} discusses the main features, requirements, and enabling technologies for greening WET-enabled networks. Moreover, \cite{9770091} propose a WET-enabled architecture for realizing sustainable smart agriculture applications, and  \cite{9319211} discusses variants for integrating renewable energy to power PBs. Recently, the authors in \cite{lopez.2023} have addressed the development of WET from the safety perspective for high-power applications. Therein, the authors overview RF electromagnetic field (RF-EMF) exposure metrics and safety limits and discuss approaches to realize safety-aware protocols. In addition, they also discuss low-power implementations for multi-antenna WET systems. Finally, recent works, such as \cite{8839968}, have identified security threats that can result in energy outages in WET-enabled networks and for that have proposed blockchain-based solutions. 

All in all, the aforementioned works have only focused on greening the operation phase of WET. However, social and economic aspects and thorough environmental impact assessments have not been jointly discussed. Indeed, a comprehensive evaluation of the environmental consequences of WET requires an accurate life cycle assessment of the technology. Notice that such a study would encompass all the stages in a product's lifetime, including the extraction and processing of raw materials, manufacturing and distribution, operation, and ultimately, the recycling or final disposal of the constituent materials \cite{portilla2023wirelessly}. From such a study, one can derive the carbon footprint of a product that can be used to label the product itself, devise better carbon trading strategies, and promote manufacturing processes with lower carbon footprint \cite{6742619}.

\begin{figure}[t!]
    \centering
    \includegraphics[width=\columnwidth]{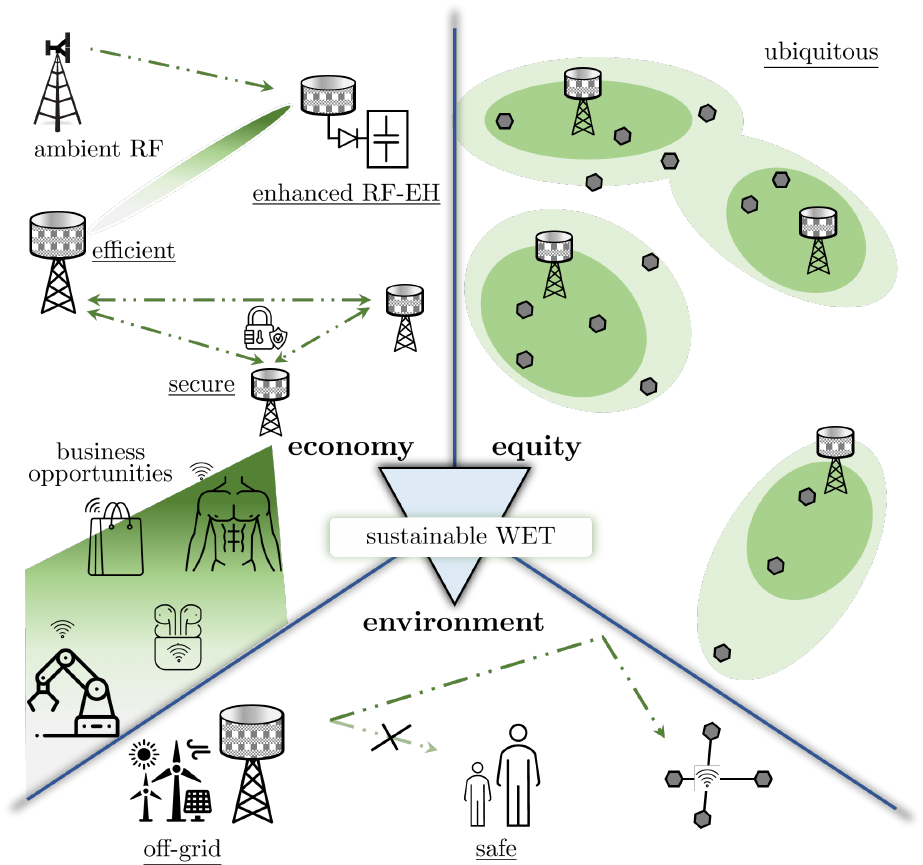}
    \caption{Towards realizing sustainable WET.}
    \label{fig:sustainableWET}
\end{figure}

As illustrated in Fig.~\ref{fig:sustainableWET}, herein we define sustainable WET as the synergy among economic prosperity, social equity, and environmental health to cope with current and future sustainable IoT quality-of-service (QoS) charging goals, including minimum wastage of resources. Sustainable WET strategies are fundamental for enabling new use cases such as zero-energy devices, extreme edge devices, and massive network deployments \cite{alves2021wireless,9269936}; thus, promoting new business opportunities. It also promotes technological innovation at both sides of the WET link to efficiently use renewable sources during the product's lifecycle. Besides maximizing the end-to-end conversion efficiency, sustainable WET also considers how to: i) efficiently power the PBs with eco-friendly energy sources, ii) secure the energy transactions (e.g., PB-to-PB and PB-to-IoT devices), iii) enable a flexible deployment of the PBs, iv) minimize the overall expenses of WET-enabled networks, and v) provide a ubiquitous charging service with compliant levels of RF pollution. This implies a full disconnection of WET from the fuel-based grid and the adoption of low-complexity, safe, and energy-efficient charging strategies, while avoiding security breaches that compromise the potentially limited energy budget. Besides, sustainable WET demands a proper design of the RF-EH circuitry as it also determines the overall conversion efficiency and reliability of the system. 

This article discusses challenges for enabling sustainable WET and possible solutions. Our contributions are sixth-fold:
\begin{itemize}
    \item We define and discuss sustainable WET for supporting IoT scalability. Moreover, we discuss technological enablers to realize the key pillars of sustainable WET indicated in the introduction.
    \item We illustrate how sustainable WET reduces overall costs compared to traditional WET and battery-powered IoT deployments via a toy numerical example. To establish a consistent ground for comparison, we propose scaling the overall expenses according to the cost per kWh of energy consumed by the PBs' deployments.
    \item We discuss the impact of ambient energy sources on the optimal deployment of a green wireless charging network. A numerical toy example shows that the optimal PBs' positioning solution trades off the energy harvested at the PBs with that harvested at the IoT devices when employing omnidirectional antennas.
    \item We discuss technological enhancement in RF-EH receivers. We first show the impact of ultra-dense networks in making RF-EH practically appealing. Then, we elaborate on RF-EH configurations operating over multiple frequency bands and input power levels. Finally, we also discuss the benefits of multi-antenna receivers and present a numerical example comparing the performance of two different architectures for ambient RF-EH.
    \item We highlight the benefits of low-complexity WET strategies to achieve cost-effective solutions. Then, we present a toy example of optimizing the number of RF chains required by a PB to meet the charging demands of an IoT deployment.
    \item We identify challenges and research directions toward realizing our vision. We also elaborate on the possible requirements of each candidate solution.
\end{itemize} 

Table~\ref{tab:researchWET} summarizes the main differences between our proposal and the discussed state-of-the-art.

\begingroup
\begin{table*}[t!]
\caption{Related works and projection into the sustainability dimensions.}
\vspace{-1mm}
\centering      
\setlength{\tabcolsep}{4pt} 
\renewcommand{\arraystretch}{1} 
\begin{tabular}{l l l c c c}
    \toprule
    Ref. & Scope & Enablers & \multicolumn{3}{c}{Sustainability dimension} \\
    & & & economy & equity & environment \\
    \midrule
    \cite{alves2021wireless} & Efficient WET & PBs' deployment, efficient beamforming, enhanced RF-EH designs & & \checkmark & \checkmark \\
    \cite{9319211} & Massive WET & PBs' (devices') architecture \& deployment, programmable medium, & & \checkmark & \checkmark \\
    & & resource scheduling, distributed ledger technology & & & \\
    \cite{8839968} & Security & Blockchain, contract theory, lightweight consensus protocol & \checkmark & & \\ 
    \cite{8664000} & Green IoT & Ambient EH, WET, wired energy trading & & & \checkmark \\ 
    \cite{9802635} & Green WET & Ambient EH, green PBs deployment & & \checkmark & \checkmark \\ 
    \cite{8309564} & Green WET & Green PBs deployment, millimeter wave communications & & & \checkmark \\ 
    \cite{8742610} & Efficient WET & Concurrent ambient EH and WET to charge IoT devices & \checkmark & & \\
    \cite{9999300} & Green WET & Green PBs deployment, dynamic PBs-to-IoT devices association & & \checkmark & \checkmark \\
    \cite{9920166} & Efficient WET & Directional antennas, dynamic PBs-to-IoT devices association & \checkmark & & \\
    \cite{10057057} & Efficient WET & Trajectory planning of mobile PB & \checkmark & \checkmark & \\
    \cite{10081057} & Efficient WET & Trajectory planning of mobile PB, directional antennas & \checkmark & \checkmark & \\
    \cite{9658133} & Efficient WET & Optimal deployment of PBs, directional antennas & \checkmark & \checkmark & \\
    \cite{9708417} & Efficient WET & Trajectory optimization of a flying PB, intelligent reflective surface & $\checkmark$ & $\checkmark$ & \\
    \cite{9770091} & Green WET & Distributed green energy storage, nomadic WET & $\checkmark$ & $\checkmark$ & \\
    \cite{lopez.2023} & High-power WET & Low-power PBs, EMF-compliant strategies, near-field RF charging & \checkmark & & \checkmark \\
    This work & Sustainable WET & Green WET, secure energy transactions, gPBs's flexible deployment, & \checkmark & \checkmark & \checkmark \\
    & & minimize the overall expenses, ubiquitous charging with minimum & & & \\
    & & RF pollution & & & \\
\hline
\end{tabular}
\label{tab:researchWET}         
\end{table*}
\endgroup

\section{Disconnecting WET from the grid}
Commonly, PBs rely on the grid as the primary source for powering IoT deployments \cite{8839968, 9536560}. Since the energy in the power grid usually comes from burning fossil fuels, the resulting WET is not green by default. Besides, having the PBs anchored to fixed positions limits their deployment and mobility. On the other hand, battery-powered PBs are not sustainable if batteries must be manually charged/replaced frequently, which may turn them into unreliable/limited energy sources for the IoT devices. 

The full disconnection of WET from the fuel-based grid not only has a positive impact on the environment but also encourages the use of self-sufficient PBs which can provide a ubiquitous charging service, as we discuss next.

\subsection{Green WET}\label{subsec:greenWET}
\begin{figure}[t!]
    \centering
    \includegraphics[width=\columnwidth]{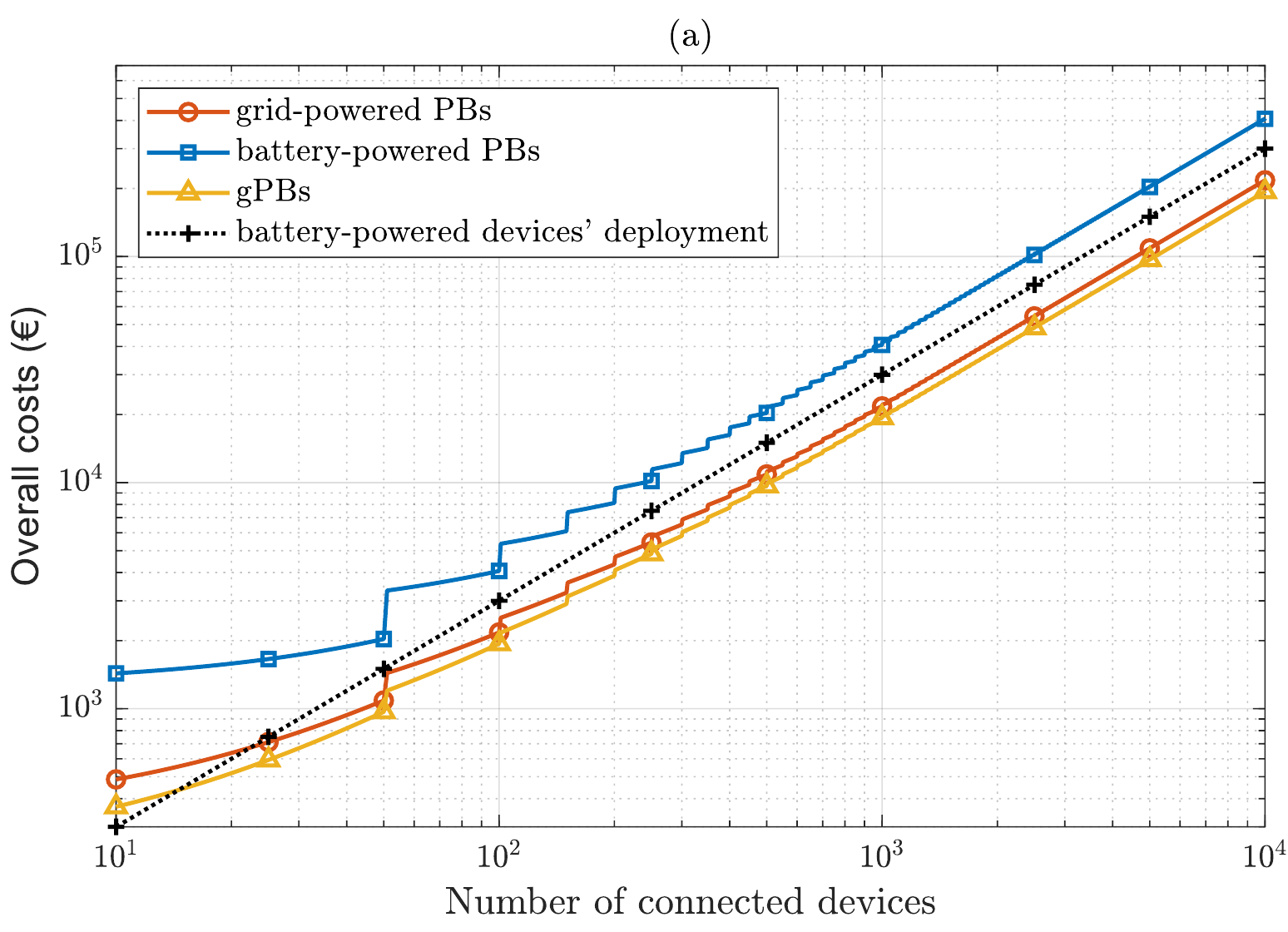}\\
    \vspace{0.5em}
    \includegraphics[width=\columnwidth]{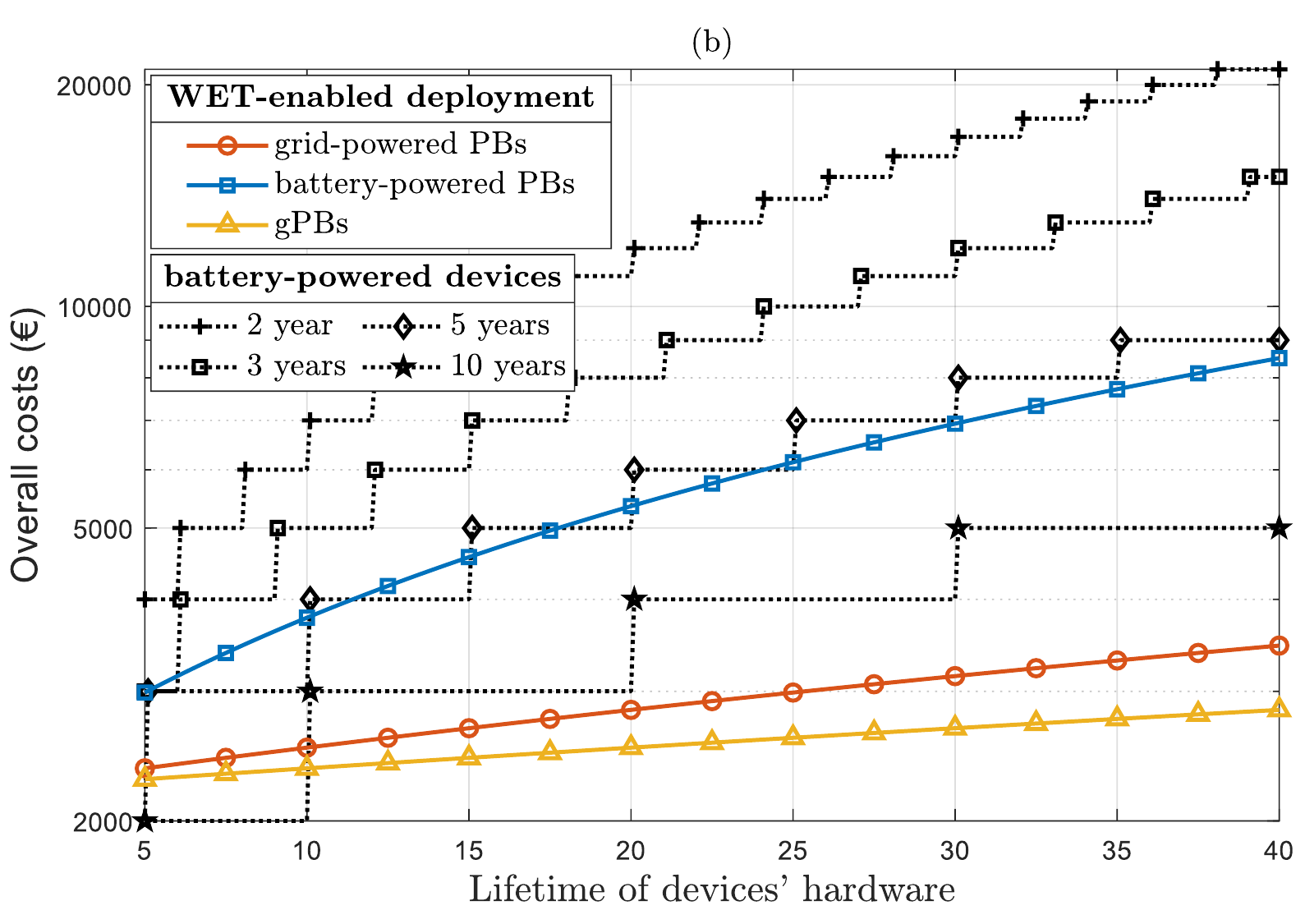}
    \vspace{-1.5em}
    \caption{Qualitative assessment of the overall costs vs: (a) number of connected devices considering $15$~years of lifetime of the IoT devices' hardware, and (b) lifetime of IoT devices' hardware for $100$ deployed devices and several battery lifetimes of the IoT devices.}
    \label{fig:OPEX}
    \vspace{-5mm}
\end{figure}

PBs have a larger form factor and better hardware/connectivity capabilities than IoT devices. This allows them to incorporate a more efficient EH circuitry that could possibly harvest from multiple ambient sources simultaneously; hence reducing the uncertainty of the total harvested energy. We will refer to the PBs powered by renewable energy sources as green-powered PBs (gPBs).

gPBs contribute to sustainable WET in the economic and environmental dimensions since the autonomous generation of electricity reduces the overall costs. To illustrate this, in Fig.~\ref{fig:OPEX}a we compare the overall costs of WET-enabled IoT deployments, including grid-powered PBs, battery-powered PBs, gPBs, and the baseline scenario where the devices rely solely on their batteries. For this example, we deploy a PB for every $50$ IoT devices. For WET-enabled scenarios, we adopt the Powercast TX91503 PowerSpot\textsuperscript{\textregistered} RF
Wireless Power transmitter\footnote{For more information please check \protect\url{https://www.powercastco.com/}} as the reference PB equipment, whose cost per unit is $\text{\texteuro}100.00$. Moreover, we scale installation and operational expenses according to the cost per kWh consumed by the PB's network in each scenario. Specifically, we adopt the values $\text{\texteuro}0.30/kWh$, $\text{\texteuro}0.15/kWh$, and $\text{\texteuro}1.50/kWh$ for the grid-powered PBs, gPBs, and battery-powered PBs, respectively. Besides, we consider that the PBs operate $24/7$ during the lifetime of the IoT devices' hardware with a power consumption of $6$~W. Unless stated otherwise, we assume that the lifetime of an IoT device's battery is five years and that maintenance accounts for $50\%$ of the installation cost of $\text{\texteuro}20.00$ per IoT device. Finally, note that the overall costs are computed over the lifetime of the IoT devices' hardware.

Note that the baseline scenario is the most cost-effective solution for powering a few devices. However, as the number of devices increases, deploying grid-powered PBs, and specially gPBs, reduces the overall costs; therefore, promoting business opportunities and enabling more use cases. For this example, we assume the devices' battery lifetime is fixed and matches the design expectations. Nevertheless, inaccurate hardware power profiling and battery imperfections may increase the overall costs far from expected as suggested by Fig.~\ref{fig:OPEX}b. We have assumed in both examples that WET is feasible for powering even the most energy-demanding IoT deployment shown in Fig.~\ref{fig:OPEX}b.

Finally, it is worth noticing that the numerical values indicated in this example are for qualitative comparison purposes and hence do not reflect exact prices in the market. Herein, we have assumed that the cost per kWh in the grid-powered PB deployment is higher than that of the gPBs' scenario, despite being more cost-effective at the generation point. This is because, in such systems, maintenance costs of the power distribution network and losses can get higher than those relying on locally generated electricity generation as in the gPBs' scenario. Finally, we set the highest operation costs for the battery-powered PBs' deployment to account for the size/quality of the energy storage system and the frequent maintenance needed to recharge the batteries.

\subsection{Ubiquitous WET}\label{subsec:ubiquitousWET}
Conveniently deploying multiple PBs is critical to eliminate blind spots in the network and distribute the energy according to the application requirements. It can also contribute to social equity by promoting ubiquitous wireless charging and enabling new use cases.

However, the optimal deployment of gPBs brings new challenges. Traditional PBs' deployment algorithms mostly focus on maximizing devices' harvested energy/rate functions and/or minimizing the energy consumption, installation costs, and/or the RF-EMF radiation of the PBs. Meanwhile, deploying gPBs adds a new degree of freedom to the corresponding optimization problems since the availability/intensity of ambient energy may influence the deployment strategies. To cope with that, the literature considers two modeling approaches of the energy arrival of ambient sources namely deterministic and stochastic. Deterministic models suit applications with predictable and slowly varying energy arrivals. For instance, it is possible to attain empirical formulas for predicting the incident energy from ambient sources using historical weather records \mbox{\cite{9411725}}. When such an approach becomes unappealing, the system can be designed for the worst-case scenario assuming bounded incident energy, which can be estimated with enough accuracy. With this, a network designer can devise a (sub-optimal) power management strategy considering the IoT device's energy budget at each time instant, the minimum incident energy, maximum energy consumption, and the battery imperfections. Another approach lies in utilizing previous measurements to feed time series analysis tools, e.g., moving average, exponential smoothing, and autoregressive integrated moving average, to realize short-term predictions of the incident energy \mbox{\cite{8968626}}, Unfortunately, this approach is not suitable for dealing with datasets with missing data, while the memory requirements and computation time to achieve low prediction errors could be prohibitive, especially for large datasets. Finally, machine learning (ML) approaches, such as the random forest algorithm, convolutional neural networks, and long short-term memory, have shown potential for capturing the spatio-temporal characteristics of the incident ambient energy. Such approaches are not tied to any physical model of the atmosphere, and thus, are simpler and require less computational resources than state-of-the-art numerical weather prediction methods \mbox{\cite{schultz2021can}}.

Meanwhile, stochastic models describe the energy arrival either as a time-correlated or uncorrelated random process. The stochastic assumption aims at mimicking the uncertainty of most ambient sources. For instance, Weibull and Gamma distributions can describe the average wind power density and solar radiation, respectively \mbox{\cite{abdulkarim2015statistical}}. When harvesting from a single source becomes insufficient, it is desirable to have a hybrid EH solution that allows a device to collect energy from different sources and potentially boost the reliability of the IoT application. One effective way to characterize the total average output power of a hybrid EH solution is by using a Gaussian mixture model (GMM) \mbox{\cite{8843930}}. With this approach, similar sources are classified into non-overlapping clusters. Then, each cluster is modeled with a Gaussian distribution. The final GMM consists of the sum of the individual distributions that model each cluster. The Kernel density estimation (KDE) method is also effective for modeling hybrid EH solutions such as RF-EH from multiple mobile service channels \mbox{\cite{8968626}}. Unlike the GMM approach, where one must specify the clusters and their locations, KDE is a non-parametric density estimation method where each data point corresponds to a cluster's center. Similarly, one can resort to stochastic geometry tools to model the position of multiple RF sources in a service area as a Poisson Point Process. This approach allows modeling the harvested energy as a random variable that depends on the spatial density of the RF sources, their transmit power, and the corresponding energy channels\footnote{We refer the reader to Section~\ref{subsec:WETmeetsRFEH} for an exemplary application of this model.}. Further, one can characterize the performance of the network by utilizing the average harvested energy on a randomly selected device, the energy coverage probability which is the percentage of the area where the EH devices harvest more than a certain number of energy units, and the distribution of the energy outage probability conditioned on the locations of the RF sources \cite{alves2021wireless}.

Next, we illustrate the application of an exemplary deterministic model for optimizing a gPBs' deployment to maximize the received RF power of the worst IoT device of the network. Besides, consider that the gPBs' harvested energy is immediately available for use and that their maximum transmit power is $1~$W. Finally, inspired by the approach adopted in \cite{8843930}, we model spatial variations of the average ambient power as a weighted sum of Gaussian functions. The Gaussian bells have been centered at $\{-5,-5\}$, $\{5,-5\}$, $\{5,5\}$, $\{-5,5\}$, and $\{0,0\}$, respectively. As per their diagonal covariance matrices, we considered that the value of non-zero entries is the same and equal to $5$, $3$, $2$, $7$, and $8$, respectively.

\begin{figure}[t!]
    \centering
    \includegraphics[width=\columnwidth]{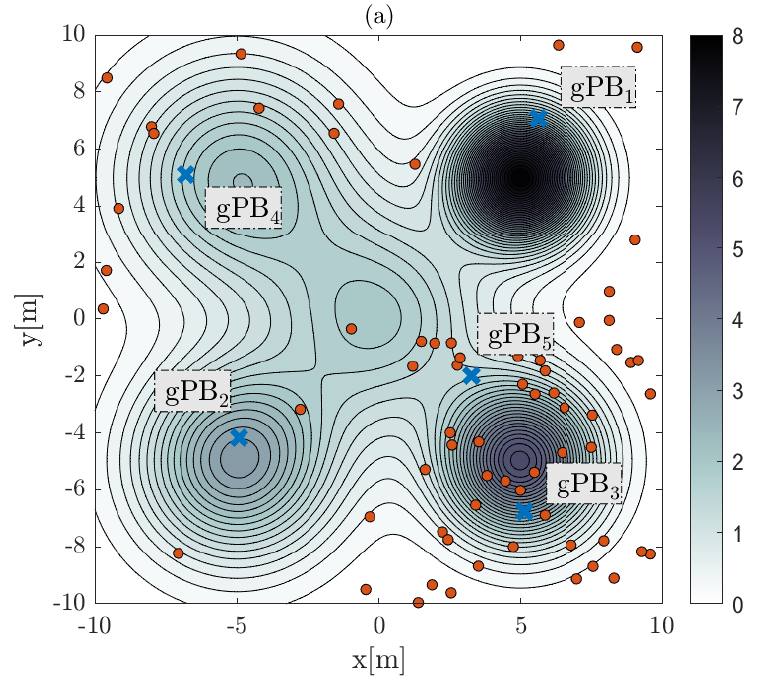}\\
    
    \includegraphics[width=\columnwidth]{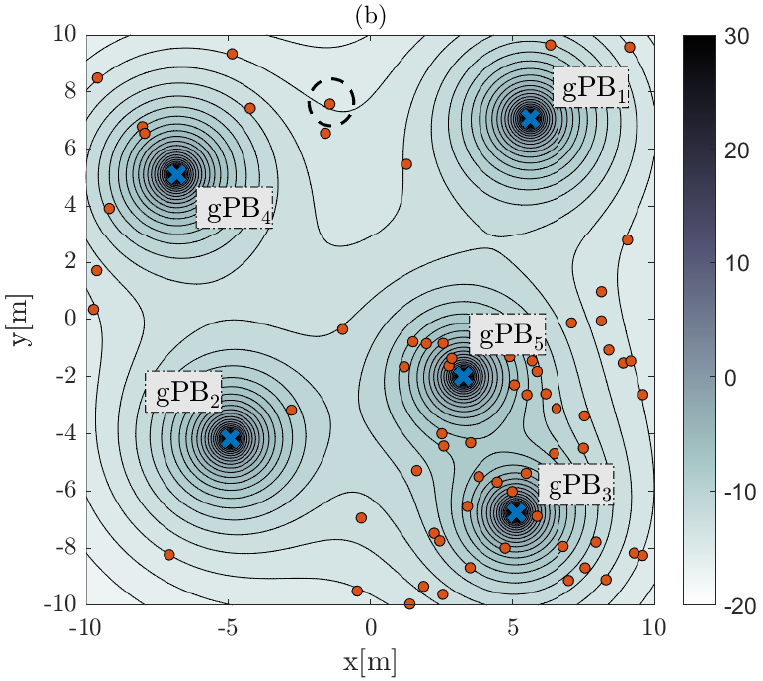}
    \vspace{-1.5em}
    \caption{gPBs' deployment (blue crosses) that maximizes the minimum average RF incident power at the IoT devices (red dots). Herein, we present the distribution of incident (a) ambient and (b) RF power (in W) to charge gPBs and IoT devices, respectively.}
    \label{fig:gPBs_positioning}
    \vspace{-4mm}
\end{figure}

Fig.~\ref{fig:gPBs_positioning} shows numerical results for the above scenario with channel losses computed according to the Friis model for an operating frequency of $1~$GHz. The level lines in Fig.~\ref{fig:gPBs_positioning}a and Fig.~\ref{fig:gPBs_positioning}b denote the average incident power of the ambient (in W) and dedicated RF sources (in dBm), respectively, at each location. Moreover, we have optimized the gPBs' deployment via a genetic algorithm solver. Notice that the available average ambient power at each gPBs' position for the resulting optimized deployment is different, e.g., $\mathrm{gPBs}_{1-5}$ receive $\{3.1, 2.5, 1.5, 3.1, 1.8\}$~W, respectively. This gPBs' deployment provides an average incident RF power at the worst-positioned device, which is enclosed in the dotted-line circle, of approximately $39~\mu$W. Hence, when using omnidirectional antennas, the gPBs must be close to the IoT devices and away from the locations where the incident ambient energy has maximum values. Moreover, depending on the spatial distribution of the IoT devices in the service area, some regions may necessitate deploying multiple gPBs close to each other. Finally, the reader can observe in Fig.~\ref{fig:gPBs_positioning}b how the distribution of RF power provides ubiquitous access to the charging service to the IoT devices by deploying more gPBs in the quadrant with the highest density of deployed devices.

\begin{figure*}[t]
	\centering
	\includegraphics[width=\linewidth]{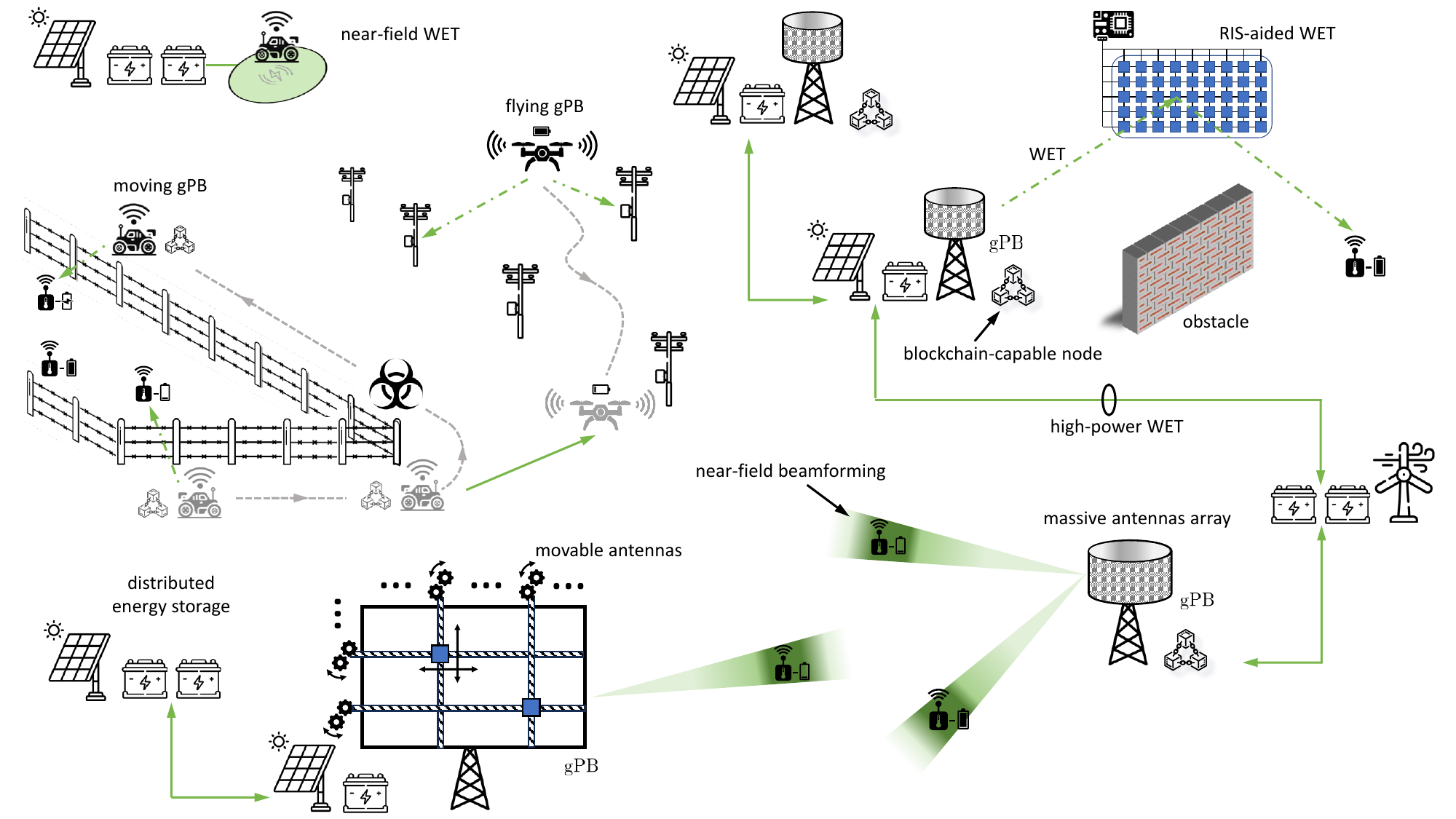}
	\caption{Promoters of sustainable WET and some use cases.}
	\vspace*{-2mm}		
	\label{fig:someWETpromoters}
	\vspace{-4mm}
\end{figure*}

For IoT applications in remote areas, and for those that require temporal communication services, deploying a network infrastructure with fixed gPBs may not be feasible. As shown in Fig.~\ref{fig:someWETpromoters}, nomadic WET, which uses moving \cite{8839968,10057057,10081057,9658133, zhu2023movable} and/or flying gPBs \cite{9708417,9770091}, can provide further flexibility to meet the service requirements under severe weather/surrounding conditions and in zones with prohibited access. In such cases, gPBs can move/fly around to power up the IoT deployments, collect measurements, and return back once the mission has been completed. Notice, though, that current regulatory frameworks for unmanned vehicles may constrain the operating parameters of such gPBs, e.g., maximum speed/altitude and minimum distance from civil infrastructure and humans.

Inevitably, under some conditions, the energy demands can far exceed the harvested energy at the gPBs, and thus compromise the WET service. In such cases, energy trading becomes an attractive solution to balance the energy available at the gPBs' network, as shown in Fig.~\ref{fig:someWETpromoters}. However, the major challenge of wireless energy trading among gPBs is that the end-to-end energy transfer efficiency may be significantly reduced. This is because the energy undergoes more transformations than in the case the devices are powered by their corresponding gPBs without recurring to energy trading. As shown in Fig.~\ref{fig:someWETpromoters}, one can ease this burden by enabling distributed energy storage systems; its constituent nodes could be seen as gPBs with larger EH circuits and energy storage capacities with the specialized role of transferring high amounts of energy to the gPBs via high-power WET links.

Finally, the advent of reflecting intelligent surfaces (RISs) in wireless systems has brought the possibility of conveniently reconfiguring the propagation environment. RISs boost the conversion efficiency by implementing a low-power reflect beamforming that guarantees a constructive interference of the reflected energy signals at the IoT devices. This helps to extend the WET coverage and avoid obstacles, as Fig.~\ref{fig:someWETpromoters} illustrates. For passive RIS, this comes at no extra energy consumption in RF chains or amplifiers, since the controller of the passive reflectors is the only active element.

\section{Enhanced RF-EH receivers}\label{sec:EnhacedRFEH}
\begin{figure}[t]
    \centering
    \includegraphics[width=.8\columnwidth]{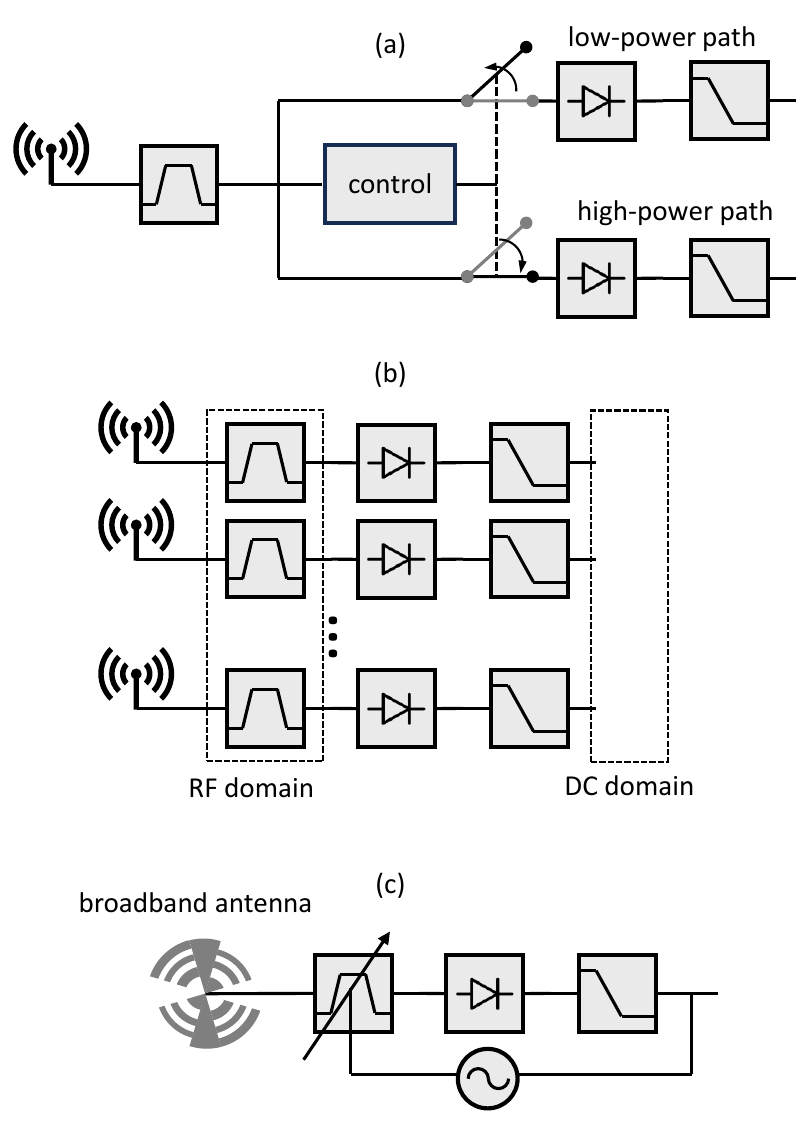}
    \caption{Generic architectures of enhanced RF-EH circuits: i) adaptive dynamic range receiver, ii) generic multiantenna receiver, and iii) broadband receiver.}
    \label{fig:enhancedEHReceivers}
\end{figure}

Engineering the RF-EH receivers to operate under different input signal conditions is key for boosting the reliability of devices' energy supply. In this section, we discuss some of the strategies to improve RF-EH from the receiver perspective. We will base our discussion on the receiver architectures illustrated in Fig.~\ref{fig:enhancedEHReceivers}.

\subsection{WET meets ambient RF-EH}\label{subsec:WETmeetsRFEH}
\begin{figure}[t]
    \centering
    \includegraphics[width=\columnwidth]{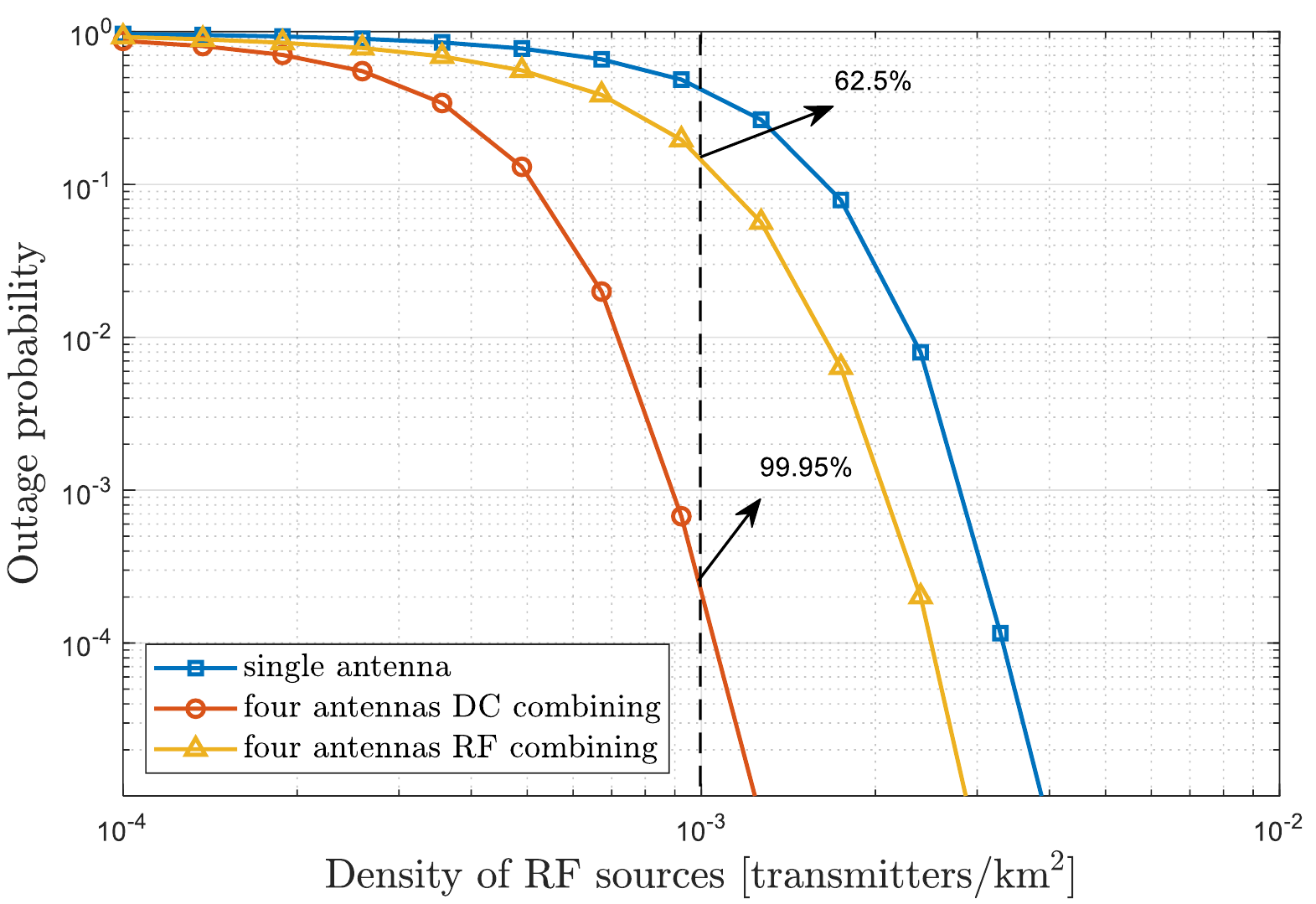}
    \caption{Outage probability of ambient RF-EH vs density of RF sources for both single and multi-antenna RF-EH circuits considering the basic architecture described in \cite{9350571}.}
    \label{fig:outageEH}
\end{figure}

Modern cities are driven by a wide range of wireless technologies that can provide average ambient RF power density levels in the order of $-25$~dBm/cm$^2$ \cite{9506895}. Typically, ambient RF-EH suits ultra-low power devices as the antenna polarization mismatch, line-of-sight (LoS) conditions, data traffic, and atmospheric conditions can weaken the incident energy. Fortunately, the advent of ultra-dense networks--- for which the estimated density is around $10^3~ \mathrm{cells}/\mathrm{km}^2$ \cite{7476821}--- will considerably reduce the propagation distance of RF signals; therefore, making RF-EH practically appealing for recycling ambient RF energy as shown in Fig.~\ref{fig:outageEH}. In this example, we adopted the sigmoidal model proposed in \cite{7264986} for the RF-EH circuit described in \cite{9350571} while illustrating the chances of not satisfying a harvesting power of $1$~mW. Moreover, we consider a $1$~W transmit power for the RF transmitters while modeling their deployment as a homogeneous Poisson point process. Besides, we consider a log-distance path loss model with exponent $2.7$ and non-distance dependent loss of $40$~dB, and channels subject to Rician fading with LoS factor $10$. We provide more details on the performance of the multi-antenna direct current (DC) and RF combining architectures in Section \ref{subsec:multiantRFEH}.

The RF-EH circuitry has the same architecture regardless of whether the RF source is dedicated or not. This allows an optimized coexistence of WET with ambient RF-EH resulting in a more reliable energy supply for the IoT devices in scenarios with high ambient power density. By enabling a dual EH mode, the IoT devices can harvest energy exclusively from ambient RF sources, and only request energy from the gPBs if needed. Conversely, ambient RF-EH can be seen as a backup or complementary energy source, if available, when WET momentarily degrades. However, notice that achieving those operating modes may require a receiver optimized for both low- and high-input power regimes, to harvest from ambient and dedicated energy transmissions, respectively. Fig.~\ref{fig:enhancedEHReceivers}a illustrates an exemplary implementation of such a receiver which has an adaptive control circuit that connects the corresponding rectifying path depending on the measured input power level.

\subsection{Advanced rectenna configurations}\label{subsec:advancedRFEHConf}
Rectennas can be designed for multiband, and extended dynamic range operation to boost the harvested energy. Multiband rectennas harvest energy at different frequency bands. As an example, one can achieve a multiband operation by connecting the output of multiple antennas, each optimized for a different frequency band, to different rectifying paths, as shown in Fig.~\ref{fig:enhancedEHReceivers}b. In general, they provide a higher conversion efficiency and more reliable operation than their narrowband counterpart. Broadband rectennas operate over a wide frequency bandwidth, by employing broadband/frequency independent antennas, and typically perform with a lower conversion efficiency than multiband rectennas. Fig.~\ref{fig:enhancedEHReceivers}c, illustrates an exemplary design of such a receiver using a tunable architecture. That is, the local oscillator circuit tunes the input-matching circuit operation to that frequency range with the highest output DC power. However, the nonlinearity of the rectifier and the matching network hinders maintaining a high performance over the operational bandwidth. Finally, extended dynamic range operation techniques aim at increasing the range of input power levels over which the power conversion efficiency of the rectifying circuit is above $20\%$. To achieve that, rectifiers must have both a high sensibility, i.e., the minimum input power level for which the DC output equals $1$~V for specified load conditions, and a low voltage drop across the diodes in the low-power regime. Meanwhile, in the high-power regime, low leakage current rectifiers are preferred to reduce power losses when the diodes are reverse-biased. A practical solution to these design requirements consists of incorporating multiple rectifying paths, each optimized for a different input power level, as shown in Fig.~\ref{fig:enhancedEHReceivers}a. Specifically, one can adaptively control the number and configuration of the rectifying blocks connected to the antenna output depending on the available power \cite{9831164}. Alternatively, one can incorporate a boosting circuit at the input to increase the voltage in the low-power regime and a self-biasing path to reduce leakage currents \cite{9531762}.

\subsection{Multi-antenna RF-EH receivers}\label{subsec:multiantRFEH}
To boost the harvested energy one can also rely on multi-antenna RF-EH receivers, whose performance improves as the number of antennas increases. In this case, the received signal is combined in the RF, or DC domain, or hybridly (Fig.~\ref{fig:enhancedEHReceivers} indicates the boundaries for these domains). RF combining aims to coherently combine the antennas' output signals to drive a rectifier circuit at the high-power operation regime. For this, CSI is needed at the IoT devices. In WET scenarios, PBs cooperation facilitates the receive CSI acquisition process, whereas an uncoordinated training period is required in ambient RF-EH, which may be energy costly \cite{9642034}. Meanwhile, DC combining adds the output of multiple rectifiers connected as one-to-one with the antenna array. Although optimal DC combining does not require CSI at the IoT devices, the fact that each rectifier operates at low input power levels makes it perform poorly with respect to RF combining, at least in WET scenarios. However, DC combining provides broader beamwidth than RF combining, since the radiation pattern seen by each individual rectenna is broader than that of the rectifier in the RF combining architecture in which the incident power is collected over the array's narrower beam. This feature allows DC combining to harvest energy from a broader range of incident directions which is beneficial in multi-source ambient RF-EH. Consequently, DC combining can outperform its RF combining counterpart in such scenarios, as shown in Fig.\ref{fig:outageEH}. It should be highlighted that this conclusion is conditioned to the presence of several strong RF sources. When the contribution of a single RF transmitter significantly dominates the received RF power, then RF combining may be preferable. In this example, we have highlighted performance improvements of these architectures for the case of $10^{-3}~\mathrm{transmitters}/\mathrm{km}^2$ with respect to the baseline single-antenna RF-EH circuit. Herein, we have neglected the power consumption for tuning the phase shifts properly in the RF combining architecture which considerably penalizes its outage probability. Besides, we have adopted the Discrete Fourier Transform -based codebook beamforming approach exemplified in \cite{9642034} for the RF combining architecture, in which the codeword yielding the highest receiving power is selected to configure the phase shifting elements. Finally, notice that multi-antenna EH receivers might challenge small form factor implementations, specially at regular microwave bands.

\section{Low-complexity, pollution-aware, and secure WET}
For gPBs, secure energy transactions and low-complexity WET strategies become crucial due to the limited energy budget. Besides, safety-aware WET strategies are key to minimize the fear to wireless and the environmental RF pollution to which traditional WET-enabled networks can significantly contribute. Next, we discuss these aspects.

\subsection{Low-complexity WET strategies}\label{subsec:lowComplexWET}
Energy beamforming (EB) techniques are beneficial for improving the end-to-end conversion efficiency without increasing the gPBs' transmit power. Unfortunately, the net benefits of EB are conditioned by the, often unaffordable, cost of estimating the instantaneous CSI at the gPBs when charging massive IoT deployments. To cope with that, data-driven approaches become appealing, as they are more resilient to imperfect CSI estimation by extracting relevant features from the acquired channel samples to adapt the EB strategies for future changes in the propagation conditions \cite{10144712}. In such scenarios, one can cast the beamforming optimization problem either into a regression problem to adjust the precoding weights or a classification problem to find the best precoder from a given codebook.

Notably, the benefits of accurate CSI-based WET strategies quickly vanish, and may even reverse, as the number of EH devices increases due to the energy-demanding training process \cite{9269936}. That is why alternative EB strategies have been proposed to rely on statistical CSI, received energy feedback, and devices' position information, which are easier to acquire and vary slowly. 

The EB's implementation also impacts the gPB's energy consumption, hardware complexity, and therefore the overall deployment cost. In this regard, analog beamforming is the simplest architecture as it operates on a single RF chain at the cost of single beam transmissions and poor spatial flexibility. Hybrid analog-digital beamforming provides further improvements at a reasonable cost \cite{7400949} by optimizing the required number of active RF chains to drive a larger number of antennas given certain QoS requirements \cite{8443446}. Let us illustrate this with an example.

\begin{figure}[t]
    \centering
    \includegraphics[width=\columnwidth]{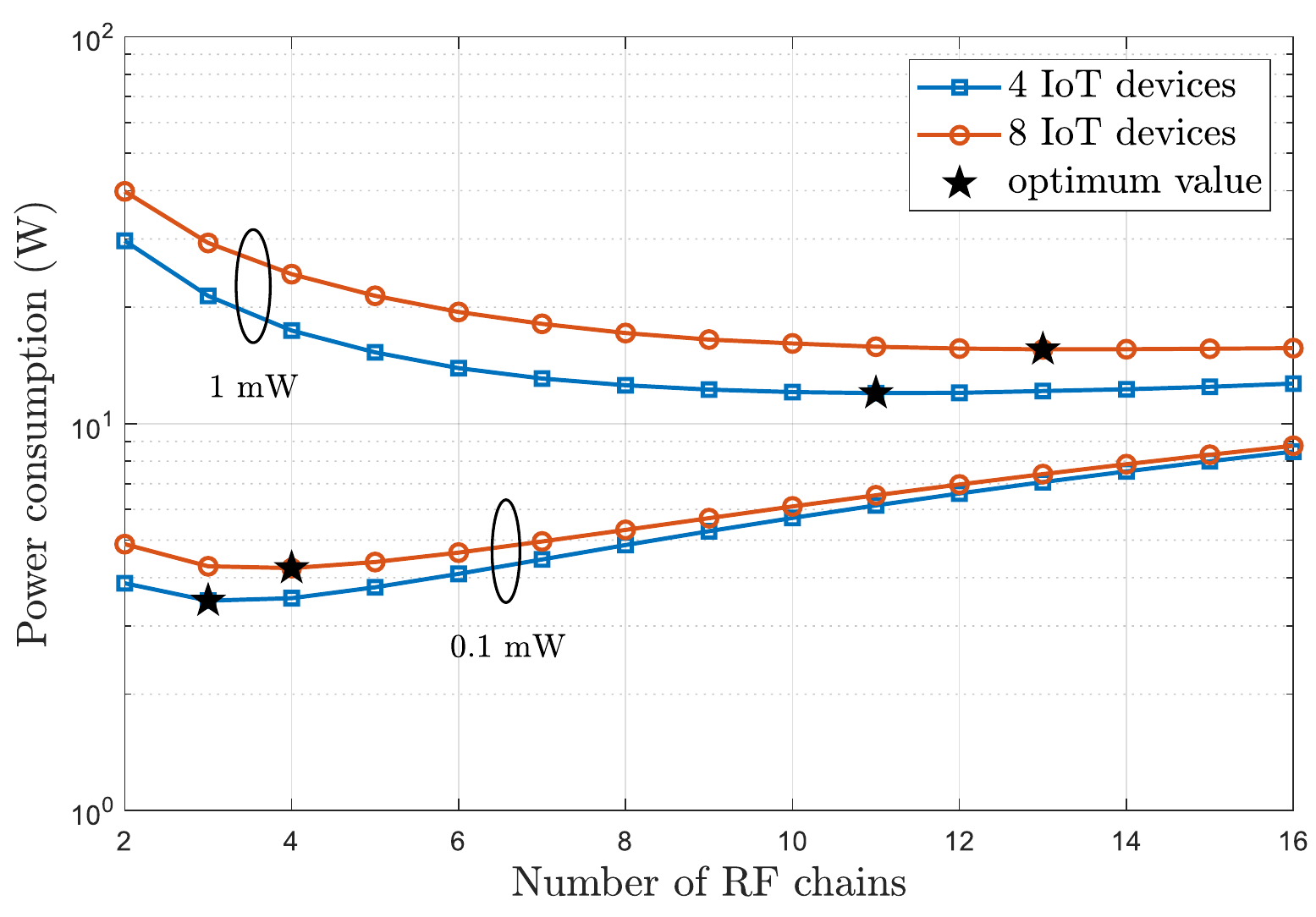}
    \caption{PB's power consumption vs the number of active RF chains for different numbers of served IoT devices and requirements on the received RF power level.}
    \label{fig:lowComplexityWET}
\end{figure}

Fig.~\ref{fig:lowComplexityWET} which shows the gPB's power consumption vs the number of RF chains when using a digital beamforming architecture, i.e., the number of antennas and RF chains is equal. Herein, the problem is to minimize the gPB's transmit power required to meet certain levels of received RF power. The served IoT devices are uniformly distributed in a circle of radius $10~$m. For this example, we have adopted the model in \cite{6503720} assuming a $35\%$ efficiency for the power amplifiers and a $0.5~$W power consumption per RF chain. This model considers that the PB's power consumption depends on the power amplifiers' efficiency, their output power, and the number of RF chains. Notice that the optimum number of RF chains increases for a larger number of deployed devices and higher requirements on the received RF power. Hence, one can design a transmit array with $M$ antennas and $M$ RF chains but then dynamically adjust the number of active RF chains to operate more efficiently in a hybrid mode.

Further energy/cost reductions can be achieved by utilizing low-resolution digital-to-analog converters \cite{9723381} and/or replacing the phase shifters by switches \cite{10144712}. Finally, recent development directions have proposed the use of lens antenna arrays \cite{10038802,6484896} to further reduce implementation complexity. This architecture resembles a hybrid analog-digital beamforming architecture where the analog beamforming part is replaced by a set of active antennas placed strategically at the focal point of a discrete set of passive lenses.

Data-driven approaches also promise to reduce the EB implementation complexity by i) approximating a known computationally demanding algorithm and ii) help designing cheaper general-purpose circuits \cite{bjornson2020two}. In the former case, the model is trained utilizing an artificial data set created by running the candidate for substitution algorithm. In the latter case, one can implement a neural network based circuit and train it during the production phase. Both applications benefit from the fact that data-driven strategies exhibit lower computational complexity in the prediction stage than model-based optimization approaches as complexity is moved from runtime to the design phase.

\subsection{Pollution-aware WET strategies}\label{subsec:pollutionAwareWET}
While increasing the incident power benefits the IoT devices, it also increases the environmental RF pollution and degrades the QoS or even causes service interruptions of nearby networks using the same spectrum. That is why investigating the impact of forthcoming extensions of the operating frequency ranges on the performance and applications of WET, may provide interesting research directions in this regard. Moreover, the exposure to high levels of RF-EMF radiation can potentially harm humans and other living species, e.g., stimulation of nerves and tissue heating for frequencies up to $10~$MHz and above $100~$kHz, respectively. The International Commission on Non-ionizing Radiation Protection \cite{international2020guidelines}, the Federal Communications Commission \cite{fcc2019guidelines}, and the IEEE \cite{8859679} dictate/recommend regulations for operation in the frequency range of $100~$kHz to $300~$GHz, for the general public and workers, and for different parts of the human body. These regulatory bodies provide a set of basic restrictions, e.g., specific absorption rate, specific absorption, absorbed power density, and absorbed energy density, that measure the absorption rate of electromagnetic energy by humans per unit of body mass or area. In practice, the basic restriction quantities are difficult to measure. Hence, reference levels in terms of incident power density, incident energy density, and electric/magnetic field strength, are also specified to provide an equivalent degree of protection as the basic restrictions.

However, the lack of conclusive studies about the long-term health effects of RF exposure at different frequencies encourages false speculations about the health effects of WET in humans. To address the fear to wireless, one can adapt the WET strategies (e.g., EB) accordingly by sensing the humans in the loop, e.g., using cameras, motion sensors, and sensing the change in ambient wireless signals caused by the presence of humans. Pollution-aware WET can also benefit from physically large antenna arrays to exploit near-field beamforming. In such a case, as shown in Fig.~\ref{fig:someWETpromoters}, the gPB focuses the energy in a particular location as opposed to steering it toward an angular direction. Finally, RIS-assisted WET can also enable the regulation of RF-EMF levels in the surroundings by limiting the exposure to non-intended users \cite{9482474}. Notice, though, that controlling the amount of RF pollution becomes challenging when using WET strategies with wide energy beams. Such cases may require frequent interruptions of the WET service and/or reduction of the transmit power levels to avoid high RF-EMF radiation levels on humans and to reduce interference.

\subsection{Secure WET strategies}\label{subsec:secureWET}
The performance of WET-enabled networks can be compromised by security attacks intended to deplete the PBs' batteries or prevent the IoT devices from being charged \cite{9170604}. Some of the attacks include repudiation of energy attacks, beamforming vector poisoning, energy state forgery, and greedy charging attacks. In repudiation of energy attacks, malicious devices deny receiving the amount of energy they agreed with the gPB. Meanwhile, beamforming vector poisoning attacks aim at corrupting the transmit strategy of the PBs. This can be accomplished by tampering with the CSI, reporting incorrect received power measurements, or jamming the energy transmissions, i.e., by causing destructive interference at the receiver end. In energy state forgery attacks, devices tamper with their battery level information. Finally, greedy devices continuously request the charging service to prevent other devices from using it. Recently, blockchain-capable PBs, as illustrated in Fig.~\ref{fig:sustainableWET}, have been proposed for securing WET-enabled networks against energy state forgery and repudiation of energy attacks \cite{8839968}. \textit{Blockchain} is a distributed ledger technology (DLT) where each transaction record is stored in a \textit{block}. In the context of energy transactions, each block can store CSI, incident RF power, devices' positions, and battery states of the served devices, just to name a few. Before adding a new block to the blockchain, a subset of gPBs must verify its authenticity by running a consensus algorithm. Once consensus has been reached, the newly created block is appended to multiple copies of the blockchain in the network. This allows every gPB to access all transaction records and also makes it difficult for a malicious entity to tamper with the information contained in the blocks. The reason is that modifying the content of one block causes its hashes to change and since each block points to the previous one using this number, the attacker must compute the hash number of subsequent blocks and convince the other blockckain participants about the authenticity of the newly created ledger.

Depending on the information stored in the blockchain, the gPBs (or a centralized entity) can detect misbehavior patterns in the network. For example, one could estimate power consumption profiles based on the amount of transferred energy over time, determine if the device is performing the activity it has signed for, validate the energy request using the device's battery capacity, and estimate the amount of received energy based on the records of the CSI and the location of the devices.

Unfortunately, as the IoT network scales up, consensus is run more frequently since gPBs must handle higher volumes of energy transactions. Notice that sustainable strategies can afford neither a computational-intensive competition, e.g., proof-of-work, nor a communication-demanding message passing, e.g., proof-of-stake, to reach consensus.

\section{Key research directions}
In this section, we discuss key research directions towards realizing sustainable WET. Table~\ref{tab:researchDirections} summarizes this discussion.

\begin{table*}[t!]
\caption{Key research directions.}
\vspace{-1mm}
\centering          
\begin{tabular}{l l l}
    \toprule
    Research direction & Candidate solution & Requirements \\
    \midrule
    Robust gPBs' deployment algorithms & Worst-case design approaches & Statistics/measurements of ambient energy \\
    & Online trajectory optimization & Vehicle's mechanical limits, government regulations, \\
    & & and accurate consumption models \\
    Intelligent and adaptive WET strategies & Stochastic control strategies & gPBs/devices' battery level, CSI statistics, \\
    & & estimations of the incident ambient energy \\
    Efficient energy trading & Hybrid wireless-wired networks & On-demand information of the trading energy cost \\
    & & among peers, available stored energy, and \\
    & & authentication protocols \\
    Environmental impact of WET & WET-oriented life cycle assessment & Trustworthy manufacturers' data from the design \\
    & & phase to hardware recycling at the application's \\
    & & end-of-cycle.\\
    End-to-end WET optimization strategies & Data-driven optimization approaches & Accurate energy consumption models, \\
    & & devices' positions information, CSI statistics, \\
    & & received energy feedback \\
    Secure energy transactions & IoT-oriented DLTs & Authentication protocols, energy efficient and  \\
    & & lightweight consensus protocols \\
    & Joint power and sensing & Trustworthy network of energy probes \\
    Performance compliant WET & Key performance indicators definition & Components' specifications, statistics of ambient\\
    & & energy, and accurate aging system's model \\
\hline
\end{tabular}
\label{tab:researchDirections}         
\end{table*}

\subsection{Robust gPBs' deployment algorithms}
Adopting a misleading model for the available ambient energy ultimately degrades the attainable ambient energy of model-based gPBs' deployment optimization algorithms. Besides, with so many variables impacting the gPBs' deployment, one may resort to efficient optimization methods to circumvent the complexity of the corresponding problem. Therefore, resilient approaches that account for model imperfections, or even the lack of a model, are key in such scenarios. One can, for instance, rely on worst-case design approaches and strategies based on the statistics of the ambient energy or live measurements. Moreover, meta-heuristic algorithms also become appealing to tackle the problem complexity as they comprise a wide family of probabilistic solvers that can achieve near-optimal solutions with high computational efficiency. Alternatively, one can rely on efficient non-convex optimization techniques that exploit the structure of the corresponding gPB's deployment problem, e.g., relaxations and convex approximations, to obtain good sub-optimal solutions.

Regarding nomadic WET implementations, one can notice that the spatio-temporal channel variations may hinder the CSI estimation process. Fortunately, air-to-ground channels are LoS-dominant with limited scattering, thus one can exploit the sparsity of the channel to device low-complexity CSI estimation schemes. Nevertheless, for the general nomadic WET channel online trajectory optimization algorithms are more appealing to deal with the channel and environmental variations \cite{9468714}. Such algorithms must also account for trajectory and charging co-design approaches considering vehicle's mechanical limits, service requirements, and regulations imposed by government authorities. For instance, altitude requirements, initial/final position, speed and acceleration limits, obstacle collision avoidance, and prohibited zones. Moreover, notice that accurate energy consumption models are necessary for these scenarios to account for the road conditions, wind direction/speed, and vehicle technology, just to name a few. Research efforts in this direction must account for these variables and have practical validation through experimental flights and measurements. 

\subsection{Intelligent and adaptive WET strategies}
When relying solely on ambient sources, gPBs must dynamically adapt their transmit strategies to save energy for periods when ambient energy is scarce or unavailable. Ideally, a gPB must have an energy neutral operation meaning that the available energy, either from the battery or the EH circuit, must meet the energy demands at any given time. Notice, though, that the available ambient energy and the energy demands of the IoT devices are difficult variables to estimate in the long term. 

Fortunately, one can rely on stochastic control theory as a potential framework for solving the optimal gPBs' power allocation under randomly-varying network dynamics. Particularly, the right tools for solving such a problem depend on the availability of the states' transition distribution, which determines whether model-free or model-based approaches can be used. For instance, dynamic programming is a model-based recursive procedure in which the main problem is decomposed into simpler sub-problems until convergence a certain convergence criterion is satisfied. Moreover, data-driven approaches have also become appealing recently for solving such problems. Particularly, reinforcement learning is a model-based/free framework for computing the best action for a given state in a dynamic system. Notice that, in the context of WET, one can model the state space of the network using gPBs/devices' batteries level, CSI statistics, measurements/estimations of the incident ambient energy.

\subsection{Efficient energy trading}
Fully wireless energy trading becomes challenging as the network may experience outages due to severe reduction in the end-to-end conversion efficiency. Hence, hybrid wireless-wired powered networks may be preferred for energy trading among fixed gPBs whereas for nomadic WET implementations highly-efficient near-field WET technologies may be more attractive.

The gPBs (or a central controller) may require the energy trading efficiency among peers, i.e., the harvested-to-spent energy ratio for each possible transaction, for the current network configuration to minimize the total losses. Notably, this information aids the trading protocol to evaluate the viability of the trading over other possible strategies, e.g., re-deploy the gPBs, considering the achievable QoS guarantees. 

\subsection{Environmental impact of WET}\label{subsec:environmImpactWET}
The potentially low end-to-end efficiency may hint to some users that WET contributes more to carbon emissions. Although this may hold for some WET implementations, such as when charging smartphones, toothbrushes, and other energy-demanding appliances, the overall environmental impact for each WET-enabled scenario is currently unclear and hence requires more studies. Noteworthy, proper key performance indicators are of paramount importance to evaluate the true carbon footprint of the available implementations. One can consider, for instance, the carbon emissions for every energy unit generated at the gPBs or transferred to the devices throughout the lifetime of the network. For that purpose, one requires, for instance, a trustworthy life cycle assessment from the design phase to the hardware recycling at the application's end-of-cycle using manufacturers' data. Moreover, the efficiency of the gPBs' EH components decays with time depending on the environmental conditions they are exposed to. For that, accurate models to consider the aging of the EH circuitry become convenient to estimate the long-term network performance, and thus the total carbon emissions during lifetime operation. Finally, one must consider that bringing eco-friendly practices to WET implementations also implicates the analysis of additional factors, such as the generated noise, land occupation, ecological damage, and mining practices for the required materials.

\subsection{End-to-end WET optimization strategies}
WET systems are composed by three fundamental components: the PB, the channel, and the RF-EH receiver. Optimizing these components independently may degrade the achievable performance since the components non-linearities couple the system performance. For instance, the non-linear response of the gPB's power amplifier may distort the amplitude and phase of the energy carrying signal, causing beamforming strategy to squint and thus reducing the harvested energy. Besides, the conversion efficiency of RF-EH receivers depends, in addition to the input power, on the waveform characteristics of the received signal, e.g., peak-to-average power ratio level, modulation, etc. Therefore, proper WET optimization strategies must target in the problem formulation the interconnection among energy-carrying signal waveform, the transmit-receive strategy, the RF-EH receiver design, and the charge/discharge characteristics of energy storage elements, just to name a few. 

Unfortunately, tractable models for the end-to-end conversion efficiency are nonexistent. That is why data-driven optimization approaches may become appealing \cite{9502719}. Moreover, accurate energy consumption models for the hardware and the network protocols, information about devices' position, CSI statistics, or receive energy feedback are also needed to carry out the optimization strategy.

\subsection{Secure energy transactions} 
Computational, communication, and energy demands of current DLTs grow exponentially with the number of deployed IoT devices. Therefore, lightweight and scalable DLTs become essential in this context. Special attention must be paid to the energy efficiency and complexity of the consensus protocols, and the design of the data structure holding each DLTs' block. Therefore, IoT-oriented DLTs that account for the limited energy and computation resources at both the gPBs and IoT devices become necessary.

For urgent transactions, gPBs can alternatively map the available energy in the network using trustworthy information sources, thus avoiding unnecessary queries to the DLTs. For that purpose, a joint power and sensing paradigm, in which a trustworthy network of energy probes continuously updates the available energy to the gPBs, may become appealing. 

\subsection{Performance compliant WET}\label{subsec:QoSGuaranteesWET}
The unpredictability of ambient energy arrivals can challenge the realization of autonomously operated gPBs. Therefore, introducing new key performance indicators becomes necessary for evaluating the performance of sustainable WET-enabled networks. For instance, the energy conversion efficiency of EH systems provides insights into the maximum harvested energy for a certain amount of input ambient energy, load impedance, transducer's quality and orientation, and environmental conditions, just to name a few. Moreover, since the available ambient energy changes randomly over time and space, one can resort to energy outage probability metrics to evaluate the network performance. In this regard, one can rely on i) the probability that the gPBs' energy budget depletes; ii) the probability of the received ambient energy being below a threshold for a certain period of time; and iii) the probability of an energy transfer operation to become practically infeasible, e.g., due to the very low conversion efficiency of the entire energy path.

The performance of the multiple components added to the gPBs, e.g., transducers, batteries, adjustable antennas, electric motors, and power management units, to sustain their operations decays at different rates depending on the environmental conditions. Besides, the fact that each component ages differently may significantly compromise the overall lifetime of sustainable WET-enabled networks. In this regard, a performance decay model of the system based on the individual components' performance will become convenient. From that one can derive the system's lifespan for a minimum accepted performance threshold, failure rate, availability, and maintenance frequency, just to name a few.

\section{Conclusion}
We introduced the concept of sustainable WET for supporting a reliable and perpetual operation of future low-power IoT deployments with minimum carbon footprint and compliant levels of RF pollution. Results evinced that deploying gPBs can reduce overall costs compared with traditional WET and battery-aided IoT deployments. We discussed different models for describing the availability of ambient energy sources, and their relevance for achieving optimal deployments of gPBs and enabling reliable ambient RF-EH. We elaborated on different implementations for green energy transmitters and illustrated relevant use cases. We reveal insights towards implementing enhanced RF-EH circuits and low-complexity multi-antenna transmitters. We emphasized sustainable WET strategies must comply with the regulations for reducing the environmental RF pollution and robust and lightweight energy trading protocols are needed for mitigating energy leaks caused by malicious attacks. Finally, we discuss the main research directions for achieving this vision.

\textit{Reproducible research}: The simulation results can be reproduced using the Matlab code available at:
\url{https://github.com/Osmel-dev/sustainable-RF-WET}

\bibliographystyle{IEEEtran}
\bibliography{IEEEabrv,references}

\end{document}